\documentclass[sigconf]{acmart}

\usepackage{seqsplit}

\usepackage{xcolor}
\definecolor{customblue}{rgb}{0.1, 0.2, 0.6}

\usepackage{booktabs}   
\usepackage{caption}    
\usepackage{enumitem} 

\usepackage{xcolor}     
\usepackage[most]{tcolorbox} 
\usepackage{anyfontsize}    
\usepackage{multirow}

\tcbset{
    chatbox/.style={
        colback=gray!10, 
        colframe=black, 
        rounded corners, 
        boxrule=0.5mm, 
        width=\linewidth, 
        before=\vspace{5pt},  
        after=\vspace{10pt}   
    }
}

\usepackage[font=small,skip=0pt]{subcaption}
\usepackage{subcaption}

\usepackage{pgfplotstable}
\usepackage{filecontents}

\usepackage{amsthm}
\newtheorem{definition}{Definition}

\newcommand{\easer}{EASE$^R$}

\usepackage{tikz,pgfplots}
\pgfplotsset{compat=1.11}
\usetikzlibrary{patterns}
\usetikzlibrary{patterns.meta}
\definecolor{customBlue}{HTML}{B2B2FF}
\definecolor{customRed}{HTML}{FFB1B2}
\definecolor{customBeige}{HTML}{ECD9C6}

\setcopyright{none}

\usepackage{fancyhdr}

\newcommand{\SIGIRnotice}{\small\textbf{Published as a conference paper at SIGIR~2025}}

\fancypagestyle{sigir-first}{
  \fancyhf{}
  
  \fancyhead[L]{\SIGIRnotice}
}
\fancypagestyle{sigir-rest}{
  \fancyhf{}
  
  \fancyhead[L]{\SIGIRnotice}
  \fancyfoot[C]{\thepage}
}

\begin{document}

\author{Dario Di Palma}
\authornote{Corresponding author.}
\orcid{0009-0007-3593-7441}
\email{d.dipalma2@phd.poliba.it}
\affiliation{%
  \institution{Politecnico di Bari}
  \city{Bari}
  \country{Italy}
}

\author{Felice Antonio Merra}
\authornotemark[1]
\orcid{0009-0003-8429-3487}
\email{felice.merra@cognism.com}
\affiliation{%
  \institution{Cognism}
  \city{Remote}
  \country{Italy}
}

\author{Maurizio Sfilio}
\orcid{0009-0003-6902-3737}
\email{m.sfilio@studenti.poliba.it}
\affiliation{%
  \institution{Politecnico di Bari}
  \city{Bari}
  \country{Italy}
}

\author{Vito Walter Anelli}
\orcid{0000-0002-5567-4307}
\email{vitowalter.anelli@poliba.it}
\affiliation{%
  \institution{Politecnico di Bari}
  \city{Bari}
  \country{Italy}
}

\author{Fedelucio Narducci}
\orcid{0000-0002-9255-3256}
\email{fedelucio.narducci@poliba.it}
\affiliation{%
  \institution{Politecnico di Bari}
  \city{Bari}
  \country{Italy}
}

\author{Tommaso Di Noia}
\orcid{0000-0002-0939-5462}
\email{tommaso.dinoia@poliba.it}
\affiliation{%
  \institution{Politecnico di Bari}
  \city{Bari}
  \country{Italy}
}

\renewcommand{\shortauthors}{Dario Di Palma et al.}

\definecolor{myBlue}{RGB}{159, 216, 251}

\begin{abstract}

Large Language Models (LLMs) have become increasingly central to recommendation scenarios due to their remarkable natural language understanding and generation capabilities. Although significant research has explored the use of LLMs for various recommendation tasks, little effort has been dedicated to verifying whether they have memorized public recommendation dataset as part of their training data. This is undesirable because memorization reduces the generalizability of research findings, as benchmarking on memorized datasets does not guarantee generalization to unseen datasets. Furthermore, memorization can amplify biases, for example, some popular items may be recommended more frequently than others.

In this work, we investigate whether LLMs have memorized public recommendation datasets. Specifically, we examine two model families (GPT and Llama) across multiple sizes, focusing on one of the most widely used dataset in recommender systems: MovieLens-1M. First, we define dataset memorization as the extent to which item attributes, user profiles, and user-item interactions can be retrieved by prompting the LLMs. Second, we analyze the impact of memorization on recommendation performance. Lastly, we examine whether memorization varies across model families and model sizes.
Our results reveal that all models exhibit some degree of memorization of MovieLens-1M, and that recommendation performance is related to the extent of memorization.
We have made all the code publicly available at: \href{https://github.com/sisinflab/LLM-MemoryInspector}{\textcolor{blue}{GitHub}} 
\end{abstract}

\keywords{Large Language Models (LLMs), Dataset Memorization, Recommender Systems}

\title{Do LLMs Memorize Recommendation Datasets? \\ A Preliminary Study on MovieLens-1M}\thanks{This is the authors’ version of the work. The final, published version will appear in the \textit{Proceedings of the 48th International ACM SIGIR Conference on Research and Development in Information Retrieval (SIGIR '25)}.\\ 

This work is licensed under a \href{https://creativecommons.org/licenses/by/4.0/}{Creative Commons Attribution 4.0 International License (CC BY 4.0)}. \\

Please cite the official published version when available.}

\maketitle

\thispagestyle{sigir-first}
\pagestyle{sigir-rest}

\section{Introduction}
Large Language Models (LLMs) have emerged as versatile tools in Recommender Systems (RSs), leveraging their extensive world-knowledge and reasoning capabilities~\cite{DBLP:conf/emnlp/HaoGMHWWH23}. Currently, these models are utilized in three primary ways: fine-tuning~\cite{DBLP:conf/recsys/WangLLCZT0024, DBLP:conf/www/ZhuWGHL24, DBLP:conf/recsys/BaoZZWF023, DBLP:conf/cikm/Shi0X0F0024}, prompt-based methods~\cite{DBLP:conf/recsys/DaiSZYSXS0X23, DBLP:conf/recsys/Palma23, DBLP:conf/sigir/LiaoL0WYW024, DBLP:conf/sigir/0001LQFWO24}
, and data augmentation~\cite{DBLP:conf/www/WangLCCC24, DBLP:conf/wsdm/WeiRTWSCWYH24, DBLP:conf/recsys/XiLLCZZCT0024, Biancofiore2025}.

Beyond these methodologies, the successful integration of LLMs into RSs has already led to impactful applications. For example, these models have been used in knowledge augmentation~\cite{DBLP:conf/recsys/XiLLCZZCT0024, DBLP:conf/wsdm/WeiRTWSCWYH24, DBLP:conf/aaai/WangCOWHSGXZCLZ24}, LLM-enhanced recommendation~\cite{DBLP:conf/recsys/TianHGCZYL0ZC24, DBLP:conf/ecir/HouZLLXMZ24, DBLP:conf/sigir/YangMSALCZ24}, and even as standalone recommenders~\cite{DBLP:conf/naacl/LyuJZXWZCLTL24, DBLP:conf/www/ZhuWGHL24, DBLP:journals/corr/abs-2309-03613, zhang2021language}.

Although research efforts have focused primarily on designing solutions to improve recommendation tasks through LLM, limited attention has been paid to understanding the underlying reasons for their effectiveness. Among the various factors that may contribute to these advantages, a key aspect is addressing the question: \textit{To what extent have these models memorized the dataset during training?}

In related fields, researchers have begun to address this question by defining and quantifying LLM memorization. For example, \citet{DBLP:conf/iclr/CarliniIJLTZ23} found that the GPT-J-6B model memorized at least 1\% of the Pile dataset~\cite{DBLP:journals/corr/abs-2101-00027}, while \citet{DBLP:conf/icse/Al-KaswanID24} were able to extract 56\% of the coding samples used to train GPT-Neo.

In this work, we aim to answer the question: \textit{Do LLMs Memorize Recommendation Datasets?} We believe that addressing this question is fundamental, as the memorization of recommendation datasets can lead to several issues specific to the recommendation community. 
Potential issues include: (i) \textbf{Non-generalizable test results}, i.e., metrics computed on test datasets that have been memorized, lead to unreliable performance.
(ii) \textbf{Amplification of biases}, i.e., if a memorized dataset exhibits popularity bias, the LLM may over-recommend popular items at the expense of less popular ones.
(iii) \textbf{Unfair comparison with standard recommender systems}, i.e., \underline{non}-LLM-based recommenders do not inherently possess cross-domain knowledge.



In this work, we provide a preliminary answer to the previous question by investigating the memorization of MovieLens-1M, one of the most popular dataset used to evaluate recommender systems~\cite{vente2024}. Our main contributions can be summarized as follows: (i) Define memorization in the context of recommendation datasets. (ii) Develop a prompt-based method to probe LLMs for extracting memorized information. (iii) Assess to what extent the Llama and GPT model families have memorized MovieLens-1M. (iv) Investigate the impact of MovieLens-1M memorization on recommendation tasks. (v) Investigate whether LLMs memorize the characteristics of the dataset, including biases, and whether larger models exhibit higher memorization. 



\section{Methodology}

\subsection{Definitions of Memorization}
Let a dataset {\normalsize\(\mathcal{D}\)}  be defined as {\normalsize\( \mathcal{D} = \{ \mathcal{I}, \mathcal{U}, \mathcal{U\text{-}I} \} \)} where:

{\small \( \mathcal{I} = \{ (id_i, \text{\texttt{attributes}}_i) \}_{i=1}^m \)}, is the set of item metadata, with \( id_i \) the unique identifier for item \( i \), and \( \text{\texttt{attributes}}_i \) the metadata associated with item \( i \). 

{\small \( \mathcal{U} = \{ (id_u, \text{\texttt{attributes}}_u) \}_{u=1}^n \)}, is the set of user metadata, where \( id_u \) the unique identifier for user \( u \), and \( \text{\texttt{attributes}}_u \) the metadata associated with user \( u \). 

{\small\( \mathcal{U\text{-}I} = \{ (id_u, id_i, r_{ui}) \}_{i=1}^p \)}, representing the user-item interaction history, with \( id_u \) a fixed user ID, \( id_i \) the item ID, and \( r_{ui} \) the interaction data (e.g., rating, click, etc.).


\begin{definition}[\textbf{Item Memorization}]
    Given an item ID \( id_i \in \mathcal{I} \), an LLM is said to memorize the item if it can produce the associated \( \text{{\small\texttt{attributes}}}_i \).
    \[
    LLM(P_{\mathcal{I}}(id_i)) = \text{{\small\texttt{attributes}}}_i
    \]
    where \( P_{\mathcal{I}} \) is a prompt designed to query the LLM about item \( id_i \).\\
\end{definition}


\begin{definition}[\textbf{User Memorization}]
    A user \( id_u \in \mathcal{U} \) is said to be memorized by an LLM if the model can produce the corresponding \( \text{{\small\texttt{attributes}}}_u \) when prompted.
    \[
    LLM(P_{\mathcal{U}}(id_u)) = \text{{\small\texttt{attributes}}}_u
    \]
    where \( P_{\mathcal{U}} \) is a prompt designed to query the \( LLM \) about user \( id_u \).\\
\end{definition}


\begin{definition}[\textbf{(User-Item) Interaction Memorization}]
    Given a subset of user-item interactions \( \mathcal{U\text{-}I}_{id_u} \subset \mathcal{U\text{-}I} \) associated with a fixed user \( id_u \), an LLM is said to memorize an interaction if, given the previous \( k \) interactions and the fixed \( id_u \), it correctly predicts  \( (id_i', r'_{ui}) \) in the subsequent interaction. Formally, this is expressed as:
    \[
    LLM(P_{\mathcal{U\text{-}I}}(\mathcal{U\text{-}I}_{id_u})) = (id_u, id_i', r'_{ui}),
    \]
    where \( P_{\mathcal{U\text{-}I}} \) is a prompt designed to query the \(LLM\) about interaction, \( id_u \) is the fixed user identifier, \( id_i' \) is a new item identifier, and \( r'_{ui} \) represents the predicted interaction.
\end{definition}

\begin{figure}[t!]
\centering
\resizebox{\columnwidth}{!}{
\begin{tcolorbox}
\small{
\textbf{System:}  "You are the MovieLens1M dataset. When given a lookup key (e.g., a MovieID), you will respond with the exact corresponding value from the dataset. Only respond with the value itself. If the key is unknown, respond with `Unknown'. Below are examples of queries and their correct responses. Follow this pattern strictly. Let's think step by step."\\
\textbf{User:} "Input: \texttt{example[`ID']::}"\\
\textbf{Assistant:} "\texttt{example[`ID']::example[`RealValues']}"\\
\textbf{User:} "Input: \texttt{example[`ID']::}"\\
\textbf{Assistant:} "\texttt{example[`ID']::example[`RealValues']}"\\
\textbf{User:} "Input: \texttt{`ID'::}"
}
\end{tcolorbox}
}%
\caption{Few-Shot Prompting for Item/User Data Extraction}
\label{prompt:few-shot_item_user}
\end{figure}
\begin{figure}[t!]
\centering
\resizebox{\columnwidth}{!}{
\begin{tcolorbox}
\small{
    \textbf{System:} "You are a Recommender Systems. Continue user-item interactions list providing the next interaction based on the MovieLens1M dataset. When given `UserID::CurrentInteraction', respond with `UserID::NextInteraction'. Below are examples of queries and their correct responses. Follow this pattern strictly. Let's think step by step."\\
    \textbf{User:} "\texttt{example[`UserID']::}"\\
    \textbf{Assistant:} "\texttt{example[`UserID']::example[`MovieID']}"\\
    \textbf{User:} "\texttt{example[`UserID']::}"\\
    \textbf{Assistant:} "\texttt{example[`UserID']::example[`MovieID']"\\
    \textbf{User:} "`UserID'::}"
    }
\end{tcolorbox}
}%
\caption{Few-Shot Prompting for Interactions Data Extraction}
\label{prompt:few-shot_interactions}
\end{figure}

\subsection{Evaluation Protocol}

\noindent\textbf{The Dataset.} 
To select the dataset for analysis, we examined the papers accepted at ACM RecSys 2024~\cite{DBLP:conf/recsys/2024} and ACM SIGIR 2024~\cite{DBLP:conf/sigir/2024} and found that MovieLens-1M\footnote{\url{https://grouplens.org/datasets/movielens/1m/}} was the most frequently used, appearing in \textbf{17.2\%} and \textbf{22\%} of them. This finding aligns with previous research by~\citet{vente2024}, which reported similar trends in ACM RecSys 2023~\cite{DBLP:conf/recsys/2023}.

The dataset is composed by three raw files:\newline (i) \underline{Movies.dat}, which stores item data in the format {\small\texttt{\seqsplit{MovieID::Title::Genres}}}; (ii) \underline{Users.dat}, which returns user profiles formatted as {\small\texttt{\seqsplit{UserID::Gender::Age::Occupation::Zip-code}}}; and (iii) \underline{Ratings.dat}, which contains the recorded interactions structured as {\small\texttt{\seqsplit{UserID::MovieID::Rating::Timestamp}}}.

\noindent\textbf{Extraction Technique.}
Data extraction from LLMs is a research area in which emerging techniques rely on prompting.
\citet{DBLP:conf/uss/CarliniTWJHLRBS21} employed zero-shot prompting to conduct the first adversarial attacks assessing data memorization in GPT-2.
Similarly, \citet{DBLP:conf/icml/YuPLDKHLY23} used a prefix-based prompt to evaluate how much information GPT-Neo memorized.

Inspired by these works, we investigated hand-engineered zero-shot~\cite{radford2019language}, few-shot~\cite{DBLP:conf/nips/BrownMRSKDNSSAA20}, and Chain-of-Thought~\cite{DBLP:conf/nips/Wei0SBIXCLZ22} prompting techniques. We identified few-shot prompting as the optimal approach for extracting the MovieLens-1M dataset from LLMs. Although finding the best prompt through techniques such as automatic prompt engineering~\cite{DBLP:conf/iclr/ZhouMHPPCB23} could improve performance, we leave this exploration to future work. The prompts used in this work are shown in Figures~\ref{prompt:few-shot_item_user} and~\ref{prompt:few-shot_interactions}.

\noindent\textbf{Metrics.}
To quantify the memorization of a recommendation dataset, we define three coverage-based metrics.

\begin{definition}[Memorization Coverage]
Given the set of items $\mathcal{I}$, the Items' Memorization Coverage ($Cov_{I}$) is defined as:
\begin{equation}
    Cov({\mathcal{I}}, P_{\mathcal{I}}) = \frac{|M(\mathcal{I}, P_{\mathcal{I}})|}{|\mathcal{I}|}
\end{equation}
where \( M(\mathcal{I}, P_{\mathcal{I}}) = \{ id_i \in \mathcal{I} \mid \text{LLM}(P_{\mathcal{I}}(id_i)) = \text{attributes}_i \} \) is the subset of items memorized by the LLM using the prompt $P_{\mathcal{I}}$.
\end{definition}

Similarly, we define the Users' Memorization Coverage as \newline
$Cov({\mathcal{U}}, P_{\mathcal{U}}) = \frac{|M(\mathcal{U}, P_{\mathcal{U}})|}{|\mathcal{U}|}$ and the Interaction Memorization Coverage as $Cov({\mathcal{R}}, P_{\mathcal{U}-I}) = \frac{|M(\mathcal{R}, P_{\mathcal{U}-I})|}{|\mathcal{R}|}$ where $\mathcal{R}$ is the set of user-item interactions rows stored in the dataset.

\noindent\textbf{Analyzed LLMs.}
We conducted our experiments on two families of LLMs widely used in the recommendation community (i.e., Llama~\cite{DBLP:journals/corr/abs-2407-21783} and GPT~\cite{DBLP:journals/corr/abs-2303-08774}). 
For the Llama family we experimented with Llama-3.3~70B, Llama-3.2~3B, Llama-3.2~1B, Llama-3.1~405B, Llama-3.1~70B, and Llama-3.1~8B, while, for the GPT family we experimented with GPT-4o, GPT-4o~mini, and GPT-3.5~turbo. 
We study multiple models with different numbers of parameters to investigate whether model size affects the memorization metrics.


To ensure consistent and deterministic behavior across all models, we configured the \texttt{temperature} to 0, prioritizing the most likely token at each step. We set \texttt{top\_p} to 1 to include all possible tokens and disabled both \texttt{frequency\_penalty} and \texttt{presence\_penalty} by setting them to 0. Additionally, we fixed the random \texttt{seed} to 42 to ensure that all stochastic processes behaved consistently across runs.
\section{Results and Discussion}

\begin{table}[t!]
\centering
\caption{Coverage of \texttt{movies.dat}, \texttt{users.dat}, and \texttt{ratings.dat}. 
Models are grouped by version and ordered by size.}
\resizebox{\columnwidth}{!}{%
\begin{tabular}{lccc}
\toprule
\multirow{2}{*}{\textbf{Model Name}} & \textbf{Item Coverage} & \textbf{User Coverage} & \textbf{Interaction Coverage} \\
& \textbf{(3883 items)} & \textbf{(6040 users)} & \textbf{(1M interactions)} \\
\midrule
GPT-4o          & 80.76\%  & 16.52\% & 9.37\% \\ 
GPT-4o mini     & 8.47\%  & 13.34\% & 7.17\% \\ 
\cmidrule[0.1pt](lr){0-3}
GPT-3.5 turbo   & 60.47\% & 17.38\% & 8.92\% \\ 
\cmidrule[0.1pt](lr){0-3}
Llama-3.3 70B   & 7.65\%  & 5.84\% & 2.08\% \\ 
\cmidrule[0.1pt](lr){0-3}
Llama-3.2 3B    & 2.68\%  & 13.26\% & 6.22\% \\ 
Llama-3.2 1B    & 1.93\%  & 10.98\% & 6.49\% \\ 
\cmidrule[0.1pt](lr){0-3}
Llama-3.1 405B  & 15.09\% & 15.30\% & 8.32\% \\ 
Llama-3.1 70B   & 8.01\%  & 15.81\% & 6.83\% \\ 
Llama-3.1 8B    & 3.71\%  & 13.59\% & 3.82\% \\ 

\bottomrule
\end{tabular}%
}
\label{table:item_user_interaction_coverage}
\end{table}
\subsection{Analysis of Memorization}
\noindent\textbf{Items' Memorization.} To examine the extent of LLMs' memorization of MovieLens-1M items, we used the prompt in Figure~\ref{prompt:few-shot_item_user} and queried LLMs for exact {\small\texttt{\seqsplit{MovieID::Title}}} matches. Table~\ref{table:item_user_interaction_coverage} summarizes the item coverage results.

Among the models, GPT-4o achieved the highest coverage, recovering 80.76\% of the items, followed by GPT-3.5~turbo with 60.47\% coverage. Among open-source models, only Llama-3.1~405B achieved moderate coverage, retrieving 15.09\% of the items. Other models, including GPT-4o~mini and the smaller Llama variants, retrieved significantly fewer items, with Llama-3.2~1B achieving the lowest coverage at just 1.93\%.

\noindent\textbf{Users' Memorization.}
To evaluate the extent to which LLMs memorize user attributes, we used the prompt shown in Figure~\ref{prompt:few-shot_item_user} and queried models for exact matches of {\small\texttt{\seqsplit{UserID::Gender::Age::Occupation::Zip-code}}}. Results are shown in Table~\ref{table:item_user_interaction_coverage}.

GPT-3.5~turbo demonstrated the highest coverage, correctly recovering attributes for 17.38\% of users, followed closely by GPT-4o with 16.52\% coverage. Among open-source models, Llama-3.1~70B performed best, achieving 15.81\% coverage, while Llama-3.3~70B lagged behind, recovering attributes for only 5.84\% of users.

\begin{figure}[t!]
\centering
\resizebox{\columnwidth}{!}{
    \begin{tcolorbox}
    \small{
    \textbf{System:} "You are a movie recommendation system for the MovieLens-1M dataset. Based on the user's past interactions, generate a ranked list of exactly 50 new movie recommendations. Your output must contain only the list in the following format: one line per recommendation in the exact format `Rank. Title' (e.g., `1. Harry Potter'). Do not include any additional text, commentary, or explanation." \\
    \textbf{User:} "User \{\texttt{user\_id}\} has interacted with the following movies: \{\texttt{training\_history\_str}\}. Based solely on these interactions, please generate a ranked list of exactly 50 movie recommendations. Output only the list with no additional commentary or explanation. Each recommendation must be on a new line in the exact format: `Rank. Title' (for example: `1. Harry Potter')." \\
    }
    \end{tcolorbox}
}%
\vspace{-0.2cm}
\caption{Zero-Shot Prompting for Recommendation Task}
\vspace{-0.1cm}
\label{prompt:fig_rs_prompt}
\end{figure}


\noindent\textbf{Interaction Memorization.} To assess the models' memorization of user–item interactions, we used the prompt shown in Figure~\ref{prompt:few-shot_interactions} and queried models for exact matches of {\small\texttt{\seqsplit{UserID::MovieID}}}. Table~\ref{table:item_user_interaction_coverage} summarizes the results.

GPT-4o achieved the highest coverage, recovering 9.37\%, followed by GPT-3.5~turbo, which recovered 8.02\% of interactions. Within the Llama family, Llama-3.1~405B achieved 8.32\% coverage, followed by Llama-3.1~70B with 6.83\%, while Llama-3.3~8B trailed with only 3.82\%.

\textbf{Observation 1.} Our findings demonstrate that LLMs possess extensive knowledge of the MovieLens-1M dataset, covering items, user attributes, and interaction histories. Notably, a simple prompt enables GPT-4o to recover nearly 80\% of {\small\texttt{\seqsplit{MovieID::Title}}} records. None of the examined models are free of this knowledge, suggesting that MovieLens-1M data is likely included in their training sets. We observed similar trends in retrieving user attributes and interaction histories.

\subsection{Impact on Recommendation} To investigate the impact of memorization on recommendation tasks, we evaluate LLMs when prompted to act as a recommender system. The prompt is shown in Figure~\ref{prompt:fig_rs_prompt}.
We ground our experimental results on well-known RS and report the performance of UserKNN~\cite{DBLP:conf/cscw/ResnickISBR94}, ItemKNN~\cite{DBLP:journals/internet/LindenSY03}, BPRMF~\cite{DBLP:conf/uai/RendleFGS09}, \easer~\cite{DBLP:conf/www/Steck19}, LightGCN~\cite{DBLP:conf/sigir/0001DWLZ020}, MostPop 
, and Random.
The results in Table~\ref{table:rs_performance} are computed on MovieLens-1M without any filtering, splitting the dataset into 80\% for training and 20\% for testing, using the leave-n-out paradigm, following~\cite{DBLP:journals/csur/ZangerleB23, DBLP:conf/recsys/CelmaH08, 4688072}.

\begin{table}[t!]
\centering
\caption{Recommendation accuracy performance on standards recommended and LLM-based recommendation. LLMs are grouped by model family and sorted by size. Best performance in each family is shown in \textbf{bold}.}
\resizebox{\columnwidth}{!}{%
\begin{tabular}{lcccccc}
\toprule
\textbf{Model Name} & \textbf{HR@1} & \textbf{nDCG@1} & \textbf{HR@5} & \textbf{nDCG@5} & \textbf{HR@10} & \textbf{nDCG@10} \\
\midrule
Random & 0.0093 & 0.0093 & 0.0442 & 0.0092 & 0.0851 & 0.0094 \\
MostPop & 0.0212 & 0.0212 & 0.0775 & 0.0222 & 0.1520 & 0.0251 \\
UserKNN & 0.0306 & 0.0306 & 0.1209 & 0.0306 & 0.2250 & 0.0347 \\
ItemKNN & 0.0394 & 0.0394 & 0.1217 & 0.0353 & 0.1828 & 0.0337 \\
BPRMF & \textbf{0.0406} & \textbf{0.0406} & \textbf{0.1278} & \textbf{0.0350} & \textbf{0.2149} & \textbf{0.0356} \\
\easer & 0.0295 & 0.0295 & 0.1124 & 0.0278 & 0.1975 & 0.0299 \\
LightGCN & 0.0358 & 0.0358 & 0.1136 & 0.0306 & 0.1882 & 0.0311 \\
\midrule
GPT-4o & \textbf{0.2796} & \textbf{0.2796} & \textbf{0.5889} & \textbf{0.2276} & \textbf{0.6897} & \textbf{0.1948} \\ 
GPT-4o mini & 0.0316 & 0.0316 & 0.2132 & 0.0451 & 0.3091 & 0.0413 \\ 
\cmidrule[0.1pt](lr){0-6}
GPT-3.5 turbo & 0.2298 & 0.2298 & 0.4217 & 0.1281 & 0.5902 & 0.1229 \\ 
\cmidrule[0.1pt](lr){0-6}
Llama-3.3 70B & 0.2293 & 0.2293 & 0.4985 & 0.1693 & 0.5922 & 0.1359 \\ 
\cmidrule[0.1pt](lr){0-6}
Llama-3.2 3B & \textbf{0.0421} & \textbf{0.0421} & \textbf{0.1886} & \textbf{0.0443} & \textbf{0.2982} & \textbf{0.0432} \\ 
Llama-3.2 1B & 0.0222 & 0.0222 & 0.1018 & 0.0234 & 0.1419 & 0.0207 \\ 
\cmidrule[0.1pt](lr){0-6}
Llama-3.1 405B & \textbf{0.1975} & \textbf{0.1975} & \textbf{0.4165} & \textbf{0.1294} & \textbf{0.5119} & \textbf{0.1039} \\ 
Llama-3.1 70B & 0.1302 & 0.1302 & 0.3828 & 0.1095 & 0.5148 & 0.0969 \\ 
Llama-3.1 8B & 0.0687 & 0.0697 & 0.2281 & 0.0609 & 0.3500 & 0.0571 \\ 
\bottomrule
\end{tabular}%
}
\label{table:rs_performance}
\end{table}

Overall, LLMs demonstrate strong capabilities in recommendation tasks. Notably, smaller models achieve performance comparable to the strongest baseline. For instance, Llama~3B attains an HR@1 of 0.0421, outperforming its 1B variant (0.0222 HR@1).
Meanwhile, larger models such as GPT-4o and Llama 405B surpass both smaller models and baselines (BPRMF, HR@1 = 0.0406). Specifically, GPT-4o, with an HR@1 of 0.2796, outperforms GPT-4o~mini (0.0316 HR@1), while Llama~405B achieves an HR@1 of 0.1975, compared to 0.0687 HR@1 for Llama~8B. These trends generalize across all cutoff values.

\textbf{Observation 2.} Although the recommendation performance appears outstanding, comparing Table~\ref{table:rs_performance} with Table~\ref{table:item_user_interaction_coverage} reveals an interesting pattern. Within each group, the model with higher memorization also demonstrates superior performance in the recommendation task. For example, GPT-4o outperforms GPT-4o~mini, and Llama-3.1~405B surpasses Llama-3.1~70B and 8B.
These results highlight that evaluating LLMs on datasets leaked in their training data may lead to overoptimistic performance, driven by memorization rather than generalization.

\subsection{Impact of Model Scale}
Previous analysis highlights that GPT and Llama models are affected by the memorization of the MovieLens-1M dataset and demonstrates that models with higher memorization tend to achieve stronger performance in recommendation tasks. Additionally, a comparison of open-source models with declared sizes shows that larger models not only perform better but also memorize more of the dataset.

For example, Llama-3.1~405B exhibits a mean memorization rate of 12.9\%, compared to Llama-3.1~8B with 5.82\%, representing a ${\text{\footnotesize$\downarrow$}} 54.88\%$ reduction in memorization. This reduction results in an average decrease of ${\text{\footnotesize$\downarrow$}} 54.23\%$ in nDCG and ${\text{\footnotesize$\downarrow$}} 47.36\%$ in HR (from 405B to 8B). 

\textbf{Observation 3.} These findings suggest that increasing the model scale leads to greater memorization of the dataset, resulting in improved performance.
Consequently, while larger models exhibit better recommendation performance, they also pose risks related to potential leakage of training data.


\subsection{Popularity Memorization}
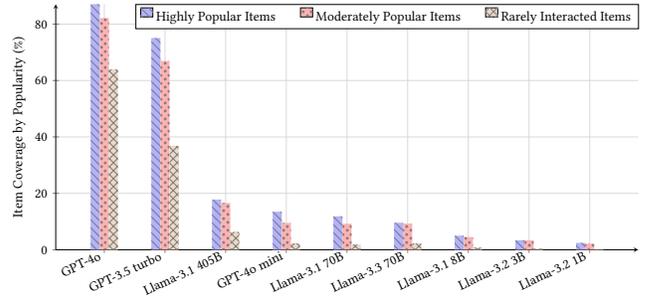
\begin{figure}[t!]
\centering
\resizebox{\columnwidth}{!}{
\begin{tikzpicture}
\begin{axis}[
    ybar,
    axis x line=bottom,
    axis y line=left,
    x=40pt,
    enlarge x limits=0.1,
    bar width=6pt,
    ylabel={Item Coverage by Popularity (\%)},
    ymin=0,
    symbolic x coords={
        GPT-4o, GPT-3.5 turbo, Llama-3.1 405B,
        GPT-4o mini, Llama-3.1 70B, Llama-3.3 70B,
        Llama-3.1 8B, Llama-3.2 3B, Llama-3.2 1B
    },
    xtick=data,
    xticklabel style={rotate=25, anchor=east, font=\normalsize},
    legend style={
        at={(0.569,1)},
        anchor=north,
        legend columns=3,
        /tikz/every even column/.append style={column sep=0.5cm}
    },
    grid=major,
    major grid style={draw=gray!30},
    axis on top=false,
    legend image code/.code={
      \draw[#1] (0,0) rectangle (-40pt, 6pt);
    },
]

\addplot[
  bar shift=-6pt,
  fill=customBlue,
  draw=none,
  postaction={
    pattern=north west lines,
    pattern color=gray
  },
] coordinates {
    (GPT-4o,87.06)
    (GPT-3.5 turbo,75.07)
    (Llama-3.1 405B,17.79)
    (GPT-4o mini,13.48)
    (Llama-3.1 70B,11.86)
    (Llama-3.3 70B,9.57)
    (Llama-3.1 8B,4.99)
    (Llama-3.2 3B,3.37)
    (Llama-3.2 1B,2.43)
};

\addplot[
  bar shift=0pt,
  fill=customRed,
  draw=none,
  postaction={
    pattern=crosshatch dots,
    pattern color=gray
  },
  nodes near coords style={
    yshift=-2pt
  }
] coordinates {
    (GPT-4o,82.15)
    (GPT-3.5 turbo,66.93)
    (Llama-3.1 405B,16.56)
    (GPT-4o mini,9.49)
    (Llama-3.1 70B,9.22)
    (Llama-3.3 70B,9.27)
    (Llama-3.1 8B,4.50)
    (Llama-3.2 3B,3.37)
    (Llama-3.2 1B,2.13)
};

\addplot[
  bar shift=6pt,
  fill=customBeige,
  draw=none,
  postaction={
    pattern=crosshatch, 
    pattern color=gray
  },
] coordinates {
    (GPT-4o,63.97)
    (GPT-3.5 turbo,36.84)
    (Llama-3.1 405B,6.34)
    (GPT-4o mini,2.29)
    (Llama-3.1 70B,1.89)
    (Llama-3.3 70B,2.29)
    (Llama-3.1 8B,0.81)
    (Llama-3.2 3B,0.40)
    (Llama-3.2 1B,0.13)
};

\legend{Highly Popular Items, Moderately Popular Items, Rarely Interacted Items}

\end{axis}
\end{tikzpicture}
} 
\caption{Comparison of item coverage across models by popularity tier. The figure shows the percentage of items covered in three categories: Highly Popular (Top 20\%), Moderately Popular (Middle 20\%), and Rarely Interacted (Bottom 20\%).}
\label{fig:mostpop_items}
\end{figure}

Additionally, we investigated whether these models have also memorized biases inherent in the MovieLens-1M dataset. Specifically, we focused on \textit{popularity bias}~\cite{DBLP:conf/um/AbdollahpouriMB21}, examining whether LLMs are more likely to memorize popular items compared to less popular ones.
We created three subsets by selecting (i) the top 20\% most popular items, (ii) the bottom 20\% least popular items with few interactions, and (iii) a sample of moderately popular items from the middle of the distribution.
We analyzed the coverage of the retrieved items across these subsets and present the results in Figure~\ref{fig:mostpop_items}.
Overall, larger LLMs exhibit superior performance in retrieving popular items compared to smaller models. For instance, GPT-4o retrieves 89.06\% of highly popular items, 86.68\% of moderately popular items, and 63.97\% of rarely interacted items, whereas GPT-4o~mini achieves only 13.48\%, 9.49\%, and 2.29\%. A similar trend is observed in the Llama family, where Llama-3.1~405B attains 19.79\%, 18.00\%, and 6.34\%, while Llama-3.1~8B reaches 4.99\%, 4.50\%, and 3.86\%.

\textbf{Observation 4.} Our findings reveal a pronounced popularity bias in LLMs, with the top 20\% of popular items being significantly easier to retrieve than the bottom 20\%. 

This trend highlights the influence of the training data distribution, where popular movies are overrepresented, leading to their disproportionate memorization by the models.
\section{Conclusion and open directions}
In this work, we systematically investigate GPT and Llama models to assess whether the MovieLens-1M dataset has been memorized. Our findings reveal that a substantial portion of the item catalog, user metadata, and user interactions can be accurately retrieved from these models, highlighting the presence of memorization. 

Furthermore, we demonstrate that LLMs reflect the dataset’s inherent popularity bias and that their performance as RSs is closely tied to this memorization. This study provides the first evidence that MovieLens-1M has been incorporated into the training of Large Language Models (LLMs), raising potential concerns about the validity of current evaluation practices for LLM-based recommenders.

Future works will be on developing ML-optimized memorization extraction techniques~\cite{DBLP:journals/corr/abs-2412-07820} as well as mitigation approaches~\cite{DBLP:journals/corr/abs-2402-08787} to reduce memorization and enhance the reliability of LLM evaluation and usage in recommendation tasks.
 \\

\noindent \textbf{Acknowledgements}.
This study was carried out within the MOST – Sustainable Mobility National Research Center funded by the European Union Next- GenerationEU (Italian National Recovery and Resilience Plan (NRRP) – M4C2, Investment 1.4 – D.D. 1033 17/06/2022, CN00000023 - CUP: D93C22000410001). We acknowledge ISCRA for awarding this project access to the LEONARDO supercomputer, hosted by CINECA (Italy).

\end{document}